 \documentclass[11pt]{article}

 \pdfoutput=1

 \usepackage[polutonikogreek,english]{babel}

 \usepackage{textgreek} 

\usepackage{bigints}

\usepackage[round]{natbib} 

\usepackage{graphicx,siunitx,xspace}
\usepackage{amsmath}
\usepackage{amssymb}
\usepackage{hyperref}
\usepackage{enumitem}
\usepackage{color}
\usepackage{enumitem}
\usepackage{float}
\usepackage{amsfonts,bm}
\usepackage{epstopdf}
\usepackage[mathscr]{euscript}

  \usepackage[normalem]{ulem}
  \usepackage{textcomp}

\usepackage[utf8]{inputenc}
\usepackage[T1]{fontenc}
\usepackage{pifont} 

\usepackage{chemformula}

\usepackage{float}
\usepackage{ifthen}
\begin{document}


\begin{center}
{\bf \Large
An English translation of the paper of Max 
\citet[][]{Planck_Ber_deut_chem_Ges_p5_1912}}
\\ \vspace*{2mm} \hspace*{-0mm}
{\bf \Large 
``\,\underline{Über neuere thermodynamische Theorien}\,''}
\\ \vspace*{2mm} \hspace*{-0mm}
{\bf \Large 
``\,\underline{(Nernstsches Wärmetheorem und Quanten-Hypothese)}\,''}
\\ \vspace*{2mm}
{\bf \Large or: ``\,\underline{On recent thermodynamic theories}\,''}
\\ \vspace*{2mm}
{\bf \Large ``\,\underline{(Nernst heat theorem and quantum hypothesis)}\,''}
\\ \vspace*{2mm}
{\bf \Large
to provide a readable version of the German content.
}
\\ \vspace*{4mm}
{\bf \large\color{black}
Translated by Dr. Hab. Pascal Marquet 
}
\\ \vspace*{2mm}
{\bf\bf\color{black}  \large Possible contact at: 
    pascalmarquet@yahoo.com}
    \vspace*{1mm}
    \\
{\bf\bf\color{black} 
    Web Google-sites:
    \url{https://sites.google.com/view/pascal-marquet}
    \\ ArXiv: 
    \url{https://arxiv.org/find/all/1/all:+AND+pascal+marquet/0/1/0/all/0/1}
    \\ Research-Gate:
    \url{https://www.researchgate.net/profile/Pascal-Marquet/research}
}
\\ \vspace*{-2mm}
\end{center}

\hspace*{65mm} Version-1 / \today

\vspace*{-4mm} 
\begin{center}
\end{center}
\vspace*{-12mm}

\bibliographystyle{ametsoc2014}
\bibliography{Book_FAQ_Thetas_arXiv}

\begin{thebibliography}{1}
\providecommand{\natexlab}[1]{#1}
\providecommand{\url}[1]{\texttt{#1}}
\renewcommand{\UrlFont}{\rmfamily}
\providecommand{\urlprefix}{URL }
\expandafter\ifx\csname urlstyle\endcsname\relax
  \providecommand{\doi}[1]{doi:\discretionary{}{}{}#1}\else
  \providecommand{\doi}{doi:\discretionary{}{}{}\begingroup
  \urlstyle{rm}\Url}\fi
\providecommand{\eprint}[2][]{\url{#2}}

\bibitem[{Planck(1912)}]{Planck_Ber_deut_chem_Ges_p5_1912}
Planck, M., 1912: {\"Uber neuere thermodynamische Theorien. (Nernstsches
  W\"armetheorem und Quanten-Hypothese) [\,On recent thermodynamic theories.
  (Nernst heat theorem and quantum hypothesis)\,]}. \textit{Berichte der
  deutschen chemischen Gesellschaft (Reports of the German Chemical Society).
  Akademische Verlagsgesellschaft (Academic Publishing Company)},
  \textbf{45~(1)}, p.5--23,
  \urlprefix\url{https://ia800708.us.archive.org/view_archive.php?archive=/28/items/crossref-pre-1923-scholarly-works/10.1002%252Fcber.191104403133.zip&file=10.1002%252Fcber.19120450103.pdf}.

\end{thebibliography}

\vspace*{-3mm} 
\begin{center}
--------------------------------------------------- 
\end{center}
\vspace*{-3mm}

I have kept {the original text of Planck (in black)} unchanged, while sometimes including {\it\color{blue}{additional notes (in blue)}, including the subsections} that were not included in the Two-Parts (I and II) text of Max Planck.
The texts \dashuline{underlined with dashlines} were highlighted by Max Planck with separated letters (like: ``{\,... a\,d\,d\,i\,t\,i\,v\,e K\,o\,n\,s\,t\,a\,n\,t\,e\, ...\,}'' instead of ``{\,... additive Konstante ...\,''\,).
Do not hesitate to contact me in case of mistakes or any trouble in the English translation from the German text (see the copy at the end of the document).

\vspace*{-3mm}
\begin{center}
\end{center}
\vspace*{-11mm}

  \tableofcontents

\vspace*{3mm}
\begin{center}
========================================================
\end{center}
\vspace*{-2mm}

\begin{center}
\underline{\Large\bf 2. On recent thermodynamic theories}
\\ \vspace*{2mm}
\underline{\Large\bf (Nernst heat theorem and quantum hypothesis)}
\\ \vspace*{3mm}
{\large\bf by Max Planck (Lecture given on 16 December 1911}
\\ \vspace*{3mm}
{\large\bf at the German Chemical Society in Berlin)}
\end{center}
\vspace*{-1mm}




Gentlemen ! In accepting the kind invitation of your honourable Board of Directors, I am attempting to develop a number of ideas before you which are of characteristic importance for the more recent \dashuline{advances in thermodynamics}. Above all, I must ask your permission to describe to you briefly some of the main features of the development of thermodynamics to date, even at the risk of unnecessarily repeating much that is already known. For only in this way will it be possible for me to describe more clearly the points on which recent research has built and to contrast what has recently been achieved with what has been known for a long time in an objectively appropriate manner.

If one attempts to gain an overview of the achievements of thermodynamics to date, it is best to make a clear distinction between two separate \dashuline{methods of research}. One is based solely on the two main theorems of thermodynamics, dispensing with the use of any more specialised hypotheses about the nature of heat. The other seeks to gain a deeper insight into the existing laws precisely from the more specialised standpoint of the mechanical theory of heat, on the basis of appropriate atomistic ideas. The advantages and also the weaknesses of each of these two methods, which have often complemented each other in the happiest way, have been described so often and thoroughly that I need not go into this point any further here. With this in mind, in the first part of the following explanations I would like to make exclusive use of the first-mentioned method, and only in the second part will I go into the atomistic significance of the newer theories, insofar as this seems possible at present.

\section{\underline{I. {\color{blue}First part (p.6-18)}}}
\label{Section-I}
\vspace*{-1mm}

\subsection{\underline{{\color{blue} Conservation of Energy (p.6-7)}}}
\label{Subsection-I-Conservation-Energy}
\vspace*{-2mm}



Of the two main theorems of thermodynamics, which were first introduced into thermodynamics under this name by Clausius, the first expresses the \dashuline{principle of the conservation of energy}. Nowadays, this theorem seems so certain and its universal validity so self-evident that one even hears the opinion here and there that the energy principle does not actually represent a theorem of experience, but rather only a kind of definition with which every future fact of experience should ultimately be able to be reconciled, if only it were interpreted appropriately.

From a practical and scientific point of view, the question of the justification of this view may seem somewhat subtle and pointless to some, but I must nevertheless ask you to allow me to add something to your discussion, because we will have to pick up on this point later. Incidentally, it was not so long ago that a lively discussion was triggered by a special occasion - back then, when it was about the energetic interpretation of the constant, considerable heat emission of radium compounds. In otherwise very noteworthy scientific articles, one could read that the surprising new discoveries had now seriously called into question the energy principle, along with many other theories that had previously been generally accepted.
Later, when the energy principle nevertheless triumphed, it was said that this was not surprising at all, because the energy principle only had a formalistic meaning and could therefore ultimately be adapted to any fact. If the principle failed in its previous form, one only needed to introduce a suitable new form of energy, for example of a potential nature, and everything would be all right again.




What about these views? If we try to formulate the energy principle properly, we will certainly find general agreement if we say that in a body system that is closed against all external influences, the energy remains constant: 
     $$\Delta U \;=\; 0 \; ,$$ 
where $U$ means the energy of the system, and the sign $\Delta$ refers to the difference of its values in two finitely different states of the system.


\subsection{\underline{{\color{blue} The value of Energy (p.7-10)}}}
\label{Subsection-I-Value-Energy}
\vspace*{-2mm}

But of course this equation only makes sense if you are able to actually specify the value of the energy $U$ in a certain state of the system, and in this respect it is still inadequate in its present form. For if one were to ask, for example: What is the energy of 1 g of water at $0$°?, some people would probably be embarrassed about the correct answer. 
For the simple reason that this question has no specific thermodynamic meaning.
In fact, we never measure energies in nature, but always only energy differences: and the change that the energy of a system of bodies suffers in any natural process is equal to the sum of all external work $A$ performed against the system and all external heat quantities $Q$ supplied to the system, measured in a suitable unit: 
\begin{align}
 \Delta U \;=\; \sum A \;+\; \sum Q 
 \label{Eq_1}
\end{align}




This equation is also a generalisation of the above formulation of the energy principle, as it merges into it in the event that the system is not subject to any external influences $(A = 0, Q = 0)$.

If a system of bodies is brought from one particular state to another particular state by two different paths, the mechanical equivalent of the external effects is the same in both cases. In other words, the energy principle, applied to the transition of a system of bodies from one particular state to another, does not consist in the fact that the change in energy is equal to the sum of external work and external heat, but in the fact that the sum of external work and heat is independent of the type of transition.
From this it might indeed seem that those who claim that the energy principle is basically only a definition are in the right, since the last equation apparently teaches nothing more than the measurement of a change in energy. But anyone who speaks in this way fails to recognise the essential content of the equation. In fact, the equation contains much more than a rule for measuring energy. For if the left side of the equation (the energy difference $\Delta U$) is related to two specific states of the body system under consideration, whereby its value is completely determined, then the right side of the equation (the mechanical equivalent of the external effects $\sum A + \sum Q$) refers to any transition from the first to the second state, and the equation claims validity for each of these transitions.




From this it is immediately apparent that the energy principle is by no means a mere definition, but that it contains an assertion whose correctness can be confirmed or refuted in countless cases by measurements, always, but only when a body system can be transferred from one state to another in various ways.

For example, the temperature of a liquid can be increased in two different ways: either directly by supplying a certain number of calories or exclusively by friction, as in Joule's famous experiments with the rotating paddle wheels. 
In the first case (1), the work supplied is $A_1 = 0$ and $Q_1$ is the amount of heat supplied. 
In the second case (2), the work supplied $A_2$ is equal to the mechanical work lost through friction, and the heat supplied $Q_2$ is $= 0$. 
The energy principle requires that the difference between the energy of the fluid in the heated state and in the original state is: 
   $$ \Delta U \;=\; A_1 + Q_1 \;=\; A_2 + Q_2 \; , $$ 
i.e. in our case 
   $$ Q_1 \;=\; A_2 \; , $$ 
and this consideration is known to have led Joule to the calculation of the mechanical equivalent of heat.
It is obviously not at all important what idea one has about the nature of heat or about the process of heat generation by friction in detail, but only that the state produced by friction in the liquid is exactly the same as the state produced by the supply of heat. If, when varying the experiment by using different liquids, friction devices and temperatures, a different value of the mechanical equivalent of heat were to be found in a single case, which is entirely possible from the outset, the energy principle would be broken, and no interpretation, no adjustment, no subsequent addition could save it. Of course, countless other examples can be added to the one cited.




The energy equation takes on a particularly simple and important form if the two states to which the energy difference $\Delta U$ refers are chosen to be identical; because then $\Delta U = 0$, and the energy equation states that in a transformation that returns the body system under consideration to its original state, i.e. in a so-called cyclic process, 
the total sum of the work and heat supplied from outside is zero
If we look back from here to the spontaneous heat development of radioactive substances mentioned above, we realise that the validity of the energy principle could also be tested very well in these peculiar processes, but only if it were possible either to return a radioactive substance to an earlier state, or if it were possible to bring the substance from one state to another in two different ways. The first, however, is probably forever hopeless, but as far as the second is concerned, it seems quite conceivable to me that one day it might be possible to influence the course of radioactive processes in some way by external means, although it must be admitted that none of the attempts directed towards this goal, which is highly significant in principle, have so far been crowned with success.

As can be seen from the general formulation of the energy equation, there are usually many different ways to calculate the energy of a body system, and all different ways must lead to one and the same result. But no matter how you vary the treatment and the methods, you will always only get energy differences, never the amount of energy itself. 
This is why it is usually said that \dashuline{an additive constant always remains undetermined} in the value of the energy. 
The magnitude of this constant is completely irrelevant for the thermodynamic phenomena.
It must be admitted that there is something unsatisfactory about the indeterminacy inherent in the value of this physically and chemically extremely important quantity, and there are scientific purists who for this reason never want to speak of energy itself, but only of energy transformations. They refuse in principle to regard energy as a property of the body itself, arguing that the energy of a body can only ever be measured by external influences on the body and that its significance must therefore be sought outside the body. But this point of view, however justified it may initially appear, is nevertheless unfruitful:  
if applied consistently, it does not lead to new perspectives, but only to a very uncomfortable complication of the approach and calculations, without providing any advantage.
It is therefore undoubtedly advisable, if not for other reasons, then at least for practical reasons, to calculate directly with the energies of the bodies themselves, and to accept the indeterminate additive constant for the time being, since one can be sure that when calculating measurable heat tones or work outputs, the indeterminate constants still attached to the energies of the bodies involved will cancel each other out by subtraction.




However, for the sake of completeness, I must not omit to mention here that the gap thus left in the definition of energy by thermodynamics has recently been filled from a completely different angle. According to the modern Lorentz-Einstein \dashuline{principle of relativity}, the absolute amount of energy of a body at rest, if the external pressure is negligible, is in mechanical mass equal to the product of its mass in the square of the propagation speed of light in a vacuum -- a tremendously large number which, however, does not appear anywhere in thermodynamics and has therefore not yet gained any practical significance.

\subsection{\underline{{\color{blue} The second law -- Entropy (p.10-11)}}}
\label{Subsection-I-Entropy}
\vspace*{-2mm}

 

If I have gone into some detail in my remarks on the first law of thermodynamics, this was done with the intention of simplifying the discussion of the second law. This is because the considerations that I have emphasised so far can easily be transferred to the second law.



The core of the second law is, as is well known, the empirical theorem that, in short, the world is in a constant state of progress, i.e. that it is impossible to return completely to an earlier state. Therefore, not a single natural process can be completely reversed.

The mathematical formulation of this theorem is made possible by the fact that for each system of bodies a quantity can be specified which is determined by the respective state of the system and which has the peculiarity of always changing unilaterally, namely always increasing, never decreasing, in all physical and chemical changes that can take place in the system if it is closed off against all external influences. \dashuline{Clausius} called this quantity, which, like energy, is to be regarded as a certain, very specific property of the body system in question, \dashuline{the entropy $S$} of the system, and the general formulation of the second law for any change of state of an externally closed body system is therefore: 
     $$\Delta S \;>\; 0 \; , $$ 
where $\Delta S$ denotes the difference between the values of entropy in the later and earlier state.
This inequality guarantees the one-sidedness of all natural processes.
However, it also exhausts the entire content of the second law, because it provides no information about the amount of the increase in entropy. 
Only in the ideal limiting case of reversible processes, to which we will restrict ourselves in the following, does the inequality reduce to the equation 
$$ \Delta S \;=\; 0 \; , $$ 
because then the later and earlier states of the system can swap their roles.




The last equation, which is valid for all reversible processes, is linked to exactly the same considerations as the energy principle above. 
Of course, it only makes sense if the value of the entropy $S$ can actually be specified, and its necessary supplement is therefore found by generalising it to such a change of state of the body system under consideration in which work is performed against the system from outside and heat is supplied to the system. Then the entropy change of the system is 
\begin{align}
 \Delta S \;=\; \sum \; \frac{Q}{T} \; , 
 \label{Eq_2}
\end{align}
where $T$ is the absolute temperature of the amount of heat $Q$ supplied to the system.

This equation represents the generalisation of the previous one, just as equation (\ref{Eq_1}) for the energy change $\Delta U$ represents the generalisation of the equation $\Delta U = 0$ for closed systems. Therefore, it also leads to the corresponding conclusions: Above all, one must not believe that the equation for $\Delta S$ merely means a definition of the entropy change. For when the body system transitions from a certain state to another certain state, the entropy of the system undergoes a very specific change, regardless of the way in which the change of state takes place, and the entropy equation states that if the change of state is reversible, the expression $\sum Q/T$ is independent of the type of transition, e.g. that it is zero for a cyclic process. This is an assertion whose correctness can be verified by measurements.


However, as with energy, one only measures differences, not the absolute amount of entropy, and so one must also accept that: \dashuline{an additive constant remains undetermined in the value of entropy}. 
But this will not prevent us from speaking of entropy, just as of energy, as a quantity determined by the state itself.



So much for classical thermodynamics. Before we go any further, I will turn to some more specialised applications.

\subsection{\underline{{\color{blue} Isothermal-isobaric processes (p.12-13)}}}
\label{Subsection-I-Isothermal-isobaric}
\vspace*{-2mm}



For isothermal processes, which play a particularly important role in physical chemistry, the temperature $T$ is constant, and it follows from the last entropy equation (\ref{Eq_2}) using the energy equation (\ref{Eq_1}): 
$$ \Delta U \;-\; T \:.\: \Delta S \;=\; \sum \: A \; .$$


If you use the abbreviation 
    $$ U \;-\; T \: S \;=\; F \; , $$ 
you can write:
\begin{align}
 \Delta F \;=\; \sum \; A \; . 
 \label{Eq_3}
\end{align}



In isothermal reversible processes, therefore, the same applies to the external work done $\sum A$ as applies to the sum of external work and heat $\sum A + \sum Q$ in any process: it is independent of the way in which the transformation is carried out, and it is equal to the difference between the values of the quantity $F$ at the end and at the beginning of the process. This quantity $F$ therefore has the same meaning for the external, freely transformable work in the processes mentioned as the energy $U$ has for the sum of external work and heat in the general case, which is why it is also called ``\,\dashuline{free energy}\,'' according to \dashuline{Helmholtz}, in contrast to the total energy $U$. Specifically for cyclic processes, $\Delta F = 0$, i.e. $\sum A = 0$.


Finally, if we assume that the reversible process is not only isothermal but also isobaric, i.e. at constant pressure $p$, then the external work can be determined directly: 
   $$ \sum \:A \;=\; -\:p \:.\: \Delta V \; ,$$ 
where $\Delta V$ denotes the change in volume of the body system under consideration caused by the process.
Then we have: 
$$ \Delta U \;-\; T \:.\: \Delta S
   \;+\; p \:.\: \Delta V \;=\; 0 \; ,$$ 
or 
\begin{align}
 \Delta \left( U \;-\; T \: S \;+\; p \: V \right) \;=\; 0 \; , 
 \label{Eq_4}
\end{align}
which equation states that in reversible isothermal-isobaric changes of state the quantity 
$$ U \;-\; T \: S \;+\; p \: V \;=\; P $$ 
does not change at all, i.e. remains constant.



The application of this simple theorem extends in particular to cases of changes in the state of aggregation: evaporation, melting, sublimation, or the conversion of a substance into an allotropic modification, provided that the conversion is reversible, isothermal and isobaric. We can also express the theorem in such a way that, in the case of two coexisting phases of a certain substance, the quantity $P$, which is usually referred to as the thermodynamic potential, has the same value in both phases, relative to a certain mass. As soon as the expression of $P$ is known, the condition of equilibrium between two adjacent phases can be stated, and thus the most important of all questions in phase theory can be answered.

Now, as can be seen, the expression of $P$ contains the quantities $U$ and $S$, and since one additive constant is undetermined in each of these, two members remain undetermined in the value of $P$.
Classical thermodynamics knows no means of generally filling this gap. 
This is not as if this makes the meaning of the thermodynamic potential illusory, because a constant, even if its magnitude is completely undetermined, naturally had important properties before a variable. 
But we can see that the question of the equilibrium of coexisting phases is not fully answered by classical thermodynamics insofar as it is not possible for it to express the law of equilibrium by a conditional equation in which only those quantities occur that relate to measurable properties of the individual phases.

\subsection{\underline{{\color{blue} The Planck-Nernst's hypotheses (p.13-14)}}}
\label{Subsection-I-Nernst-hypothesis}
\vspace*{-2mm}



This was the state of the theory when, just six years ago, W. Nernst {\color{blue}(1906)} came forward with a new and surprising hypothesis whose significance, briefly summarised, lies in the fact that it defines in a very simple and general way the additive constant left undefined by classical thermodynamics in the expression of entropy $S$.


In order to fully appreciate the fruitfulness of this \dashuline{Nernst heat
theorem}{\color{blue}\footnote{\label{footnote_Planck_Nernst}$\:${\color{blue}Note that Planck used in this 1912 paper the definition of the ``\,third law of thermodynamics\,'' he derived and published in 1911 as a generalisation of the older ``\,heat theorem\,'' of Walter Nernst (1906), which stated that: both the derivatives $d(\Delta S)/dT$ and $d(\Delta Q)/dT$ are zero in the vicinity of $0$~K for all chemical reactions (namely that both $\Delta S$ and $\Delta Q$ have horizontal tangents at $0$~K when they are plotted against the absolute temperature $T$). 
Note also that the next experiments will limit the formulation to the ``\,more stable solid\,'' state at $0$~K (namely no longer valid for the liquid states), and with the possibility of residual entropy still existing at $0$~K like for the water Ice due to hydrogen bonds in the \ch{H2O} molecule, as explained later on by Pauling (in 1935) and Nagle (in 1966). / P. Marquet.}}}, 
I consider it best to express its content in the version which, in my opinion, is the most comprehensive and at the same time the simplest. This states that: \dashuline{the entropy of a condensed} (i.e. solid or liquid) \dashuline{chemically uniform substance has the value zero at the zero point of absolute temperature}.
At first glance, this sentence also looks like a definition.
Since thermodynamics is only ever concerned with differences in entropy, the general fixation of the additive constants appears to be irrelevant for the measurements. 
However, this assumption proves to be incorrect because a condensed substance can occur in different modifications or states of aggregation, and it is not known from the outset whether its entropy at the zero point of absolute temperature is independent of the modification and the state of aggregation, as required by Nernst's heat theorem. 
Its significance will be most clearly expressed when I now turn to the treatment of some specific examples, whereby I will first discuss the conclusions to be drawn from classical thermodynamics and then the additions to it provided by Nernst's heat theorem for better comparison.

\subsection{\underline{{\color{blue} A first application of the Planck-Nernst's hypotheses (p.14-16)}}}
\label{Subsection-I-Nernst-hypothesis-first-application}
\vspace*{-2mm}


%
To calculate the entropy $S$ of a condensed body, classical thermodynamics provides the relationship according to equation (\ref{Eq_2}): 
$$ S \;=\; \bigintsss \frac{C_p}{T}\:\: dT $$ 
($C_p$ is the heat capacity at constant pressure, the integration is carried out at constant pressure).



%
The upper limit of the integral is $T$, the lower limit is left undetermined by classical thermodynamics, but the Nernst heat theorem
{\it\color{blue}(see the footnote \ref{footnote_Planck_Nernst})} 
requires zero as the lower limit, so that the entropy is completely  equal to:
\begin{align}
 S\,{\color{blue}(T)} \;=\; 
 \bigintsss_{\,0}^{\,T} \frac{C_p{\color{blue}(t)}}{t}\:\: dt 
 \;\;\;\;{\color{blue} \left[ \:+\; S_0(0)\;=\;0 \;\right] } \; . 
 \label{Eq_5}
\end{align}
\vspace*{-6mm}
{\color{blue} 
\begin{quote}
Note that the Planck-Nernst hypothesis (known as the Third law of thermodynamics) is summarized by what I have written $\:S_0(0)=0\:$ (i.e. ``\,zero as the lower limit\,''\,) in (\ref{Eq_5}).
/ P. Marquet.
\end{quote}
}



From this equation {\it\color{blue} (where I have replaced $T$ by $t$ in the integral / P. Marquet)} it follows above all that $C_p$ vanishes for {\it\color{blue}($\,t=0$~K or)} $T = 0${\it\color{blue}~(K)}, and thus the first far-reaching conclusion is that the heat capacities of all condensed substances converge towards zero with decreasing temperature, a statement that would have seemed quite strange ten years ago, but which has recently been confirmed in a striking way, particularly by the measurements of Nernst and his students.



%
Closely related to this is the other general theorem that the coefficient of thermal expansion of any condensed body disappears at absolute zero temperature. Even if this theorem has not been proven as clearly by previous measurements as the previous one, there are already certain indications of its general validity.



Let us now consider the change of state of aggregation, the transformation of two allotropic modifications of a condensed body. For the equilibrium of the two adjacent phases, classical thermodynamics provides the relationship (\ref{Eq_4}), or, if we denote the second phase by primes: 
\begin{align}
 (\,U'\:-\:U\,) \;-\; T \: (\,S' \:-\: S\,) 
 \;+\; p \: (\,V' \:-\: V\,) \;=\; 0 \; .
 \label{Eq_6}
\end{align}



Here $(U'-U) + p \:(V'-V)$ is equal to the heat of transformation $r$. Thus, using (\ref{Eq_5}) we obtain the condition of equilibrium of the two phases:
\begin{align}
 r{\,\color{blue}(T)} \;-\; T \:.
 \bigintsss_{\,0}^{\,T} 
 \frac{C'_p{\color{blue}(t)} \:-\: C_p{\color{blue}(t)}}{t}\:\: dt 
 \;=\; 0 \; . 
 \label{Eq_7}
\end{align}



If the heat of transformation $r$ and the specific heats $C'_p$ and $C_p$ are known in their dependence on temperature, the melting temperature or the transformation temperature of the body can be calculated from this. Classical thermodynamics alone cannot do this, because it leaves the lower limit of the integral undetermined.



For example, the conversion of rhombic into monoclinic sulphur is, according to Broensted's measurements, approximately equal in calories: 
  $$ r \;=\; 1.57 \;+\; 1.15 \:10^{-5} \:.\: T^2 $$ 
and 
\vspace*{-2mm}
  $$ C'_p \:-\: C_p \;=\; \frac{\partial r}{\partial T}
  \;=\; 2.3 \:10^{-5} \:.\: T
  \; . $$



According to (\ref{Eq_7}) this  
results{\color{blue}$\,$\footnote{$\:${\color{blue}The value $T \approx 369.5$ was computed by Max Planck from (\ref{Eq_7}) giving  $T^2 \approx 1.57/[\,(2.3-1.15)\,10^{-5}\,] \approx 136\,522$, and thus with an impressive accuracy of 
$\:(369.5-368.4)/368.4 \approx 0.3$~\% 
for the Planck-Nernst hypothesis\,! 
In 1911, this accuracy was similar to the one of the gravitational constant $G = 6.674\,30(15)\,10^{-11}$~m${}^3$~kg${}^{-1}$~s${}^{-2}$, evaluated by Poynting in the article ``\,Gravitation\,'' in the Encyclopaedia Britannica (Eleventh Edition of 1911) to the value of $G = (6.66 \pm 0.014)\:10^{-11}$~m${}^3$~kg${}^{-1}$~s${}^{-2}$, and thus with a similar relative uncertainty of about $0.2$~\% / P. Marquet.}}}
in the transition temperature $T=369.5$, while the direct measurement gave $368.4$.



For a chemically uniform gas in the ideal state, classical thermodynamics gives the expression of entropy: 
\vspace*{-2mm}
\begin{align}
 S \;=\; n \: \left[\: 
 C_p\:\ln(T) \:-\: R\:\ln(p) \:+\: k
 \:\right] \; ,
 \label{Eq_8}
\end{align}
where $n$ is the number of moles, $p$ is the pressure, $C_p$ is the molar heat at constant pressure, $R$ is the absolute gas constant and $k$ is a constant. At the zero point of the absolute temperature, the entropy is obviously not zero, but negatively 
infinite.{\color{blue}$\,$\footnote{$\:${\color{blue}This does not mean that the relationship (\ref{Eq_8}) does not have a physical meaning. 
The meaningless infinite values obtained for $T=0$ and $p=0$ only mean that the range of validity of the relationship (\ref{Eq_8}) must limited to the hypotheses that both $C_p$ and $R$ are constant, for instance for the usual range of temperature and pressure for which $C_p$ is nearly a constant and where the perfect gas equation $p\:V=n\:R\:T$ can be considered / P. Marquet.}}}



The Nernst heat theorem 
{\it\color{blue}(see the footnote \ref{footnote_Planck_Nernst})}
further specifies this expression in that the constant $k$ has a very specific measurable value that is characteristic of all physico-chemical properties of the gas and can therefore be called the chemical constant of the gas.


As a rule, the best way to calculate $k$ is to measure \dashuline{the vapor pressure}. 
For the equilibrium between liquid and vapor, equation (\ref{Eq_6}) in conjunction with (\ref{Eq_5}) and (\ref{Eq_8}) provides the following if the molar heat $C'_p$ refers to the vaporous state: 
$$ \frac{r{\,\color{blue}(T)}}{T} 
 \;-\; C'_p\:\ln(T) \:+\: R\:\ln(p) \:-\: k
 \;+ \bigintsss_{\,0}^{\,T} 
 \frac{C_p{\color{blue}(t)}}{t}\:\: dt 
 \;=\; 0 \; . $$
\vspace*{-5mm}
{\color{blue} 
\begin{quote}
This relationship is indeed a consequence of (\ref{Eq_6}) divided by $T$ and rewritten as:
$$
 \overbrace{
 \frac{(\,U'\:-\:U\,) \;+\; p \: (\,V' \:-\: V\,)}{T}
 }^{\displaystyle(\,r/T\,)}
 \;-\; 
 \overbrace{S'}^{\displaystyle \mbox{from (\ref{Eq_8})}}
 \;+\; 
 \overbrace{S}^{\displaystyle \mbox{from (\ref{Eq_5})}}
 \;=\; 0 \; ,
$$
with the use of prime letters for $S'$ and $C'_p$ in (\ref{Eq_8})
/ P. Marquet.
\end{quote}
}


This equation gives the tension of the saturated vapor $p$ for every liquid as a function of the temperature $T$, provided that the vapor can be treated as an ideal gas. It has already been confirmed by numerous measurements and has led to the more or less approximate calculation of the constant $k$ for a series of gases and vapors. But I hope that Mr. Nernst himself will give us a more complete report on all this.



The Nernst heat theorem {\it\color{blue}(see the footnote \ref{footnote_Planck_Nernst})} seems to have far-reaching significance not only for chemically uniform substances, but also for mixtures and solutions. The entropy of a solution, even a condensed one, is admittedly not zero at zero absolute temperature, but depends in a certain way on the concentrations of the dissolved substances, but this still means that the theorem that both the heat capacity and the thermal expansion coefficient of every condensed solution disappear at zero absolute temperature remains in agreement with the experience available to date.

\subsection{\underline{{\color{blue} Further applications of the Planck-Nernst's hypotheses (p.16-17)}}}
\label{Subsection-I-Further-applications}
\vspace*{-2mm}


To demonstrate further applications, I would like to pick out a few examples. 

As is well known, classical thermodynamics provides the condition of the law of mass action for the chemical equilibrium of different types of molecules reacting with each other within a solution: 
$$  \frac{{C_1}^{\nu_1}\;{C_2}^{\nu_2}\;{C_3}^{\nu_3}\;.\:.\:.}{{C'_1}^{\nu'_1}\;{C'_2}^{\nu'_2}\;{C'_3}^{\nu'_3}\;.\:.\:.}
\;=\; K$$ 
where the concentrations $C$ correspond to the types of molecules created during the reaction, the concentrations $C'$ correspond to the types of molecules disappearing during the reaction, while the quantities $\nu$ represent the numbers of molecules involved in the reaction {\it\color{blue}(i.e. the stoichiometric coefficients of the chemical reaction: 
\ch{${\nu'_1}$\:{M'}1 + ${\nu'_2}$\:{M'}2 + ${\nu'_3}$\:{M'}3 + ... 
    ->  $\nu_1$\:M1 + $\nu_2$\:M2 + $\nu_3$\:M3 + ...}
/ P. Marquet)}. The quantity $K\,{\color{blue}(T,p)}$ is independent of the concentrations and is determined by the temperature and pressure.


Classical thermodynamics cannot say anything about the absolute value of $K$, only its dependence on temperature and pressure can be stated. Thus, the well-known van't Hoff relation applies to the dependence on temperature: 
$$  \frac{\partial \ln[\:K\,{\color{blue}(T,p)}\:]}{\partial T} 
\;=\; \frac{r\,{\color{blue}(T,p)}}{R\:T^2} \; .$$


But the Nernst heat theorem {\it\color{blue}(see the footnote \ref{footnote_Planck_Nernst})} goes further. 

For the solution that is condensed, the following applies directly {\it\color{blue}(via an integration by part, with $u\:dv=d(u\,v)-v\:du$, of the previous equation / P. Marquet)}: 
$$ R \:.\: \ln[\:K\,{\color{blue}(T,p)}\:]
{\:\color{blue}
\;=
\bigintsss_{\,0}^{\,T} \!\!
    r\,{\color{blue}(t,p)} \:.\: \frac{dt}{t^2} 
}
\;= 
\;-\; \frac{r\,{\color{blue}(T)}}{T}
\;\:+ \bigintsss_{\,0}^{\,T} \!
          \frac{\partial\,r\,{\color{blue}(t)}}{\partial\,t} 
    \;.\; \frac{dt}{t} 
\; , $$ 
so that it only needs to measure the heat of reaction $r\,{\color{blue}(T)}$ at different temperatures to fully determine the law of chemical equilibrium {\it\color{blue} (I have, again, replaced $T$ by $t$ in the integrals / P. Marquet)}. 
{\color{blue} 
\begin{quote}
Note that the last relationship is only valid with suitable properties and reference values to be valid at $T \approx 0$ and not recalled by Max Planck, namely: 
$\:\lim \, K(T)=0$\,; 
$\:\lim \, r(T)/T = 0$\,; and finite values for 
$\:\lim \, r(t)/t^2$\, and  
$\:\lim \, (1/t)\,
  [\,{\partial\,r\,{\color{blue}(t)}}/{\partial\,t}\,]$
/ P. Marquet.
\end{quote}
}

\noindent If, on the other hand, the solution is gaseous, the chemical constants $k$ of the gaseous molecule species and the pressure $p$ are also included in the expression of $K$.


The case is particularly simple for $r=0$, i.e. for a thermoneutral reaction. For this, classical thermodynamics only states that the equilibrium constant $K$ is independent of the temperature, leaving its absolute value completely undetermined. However, the Nernst heat theorem {\it\color{blue}(see the footnote \ref{footnote_Planck_Nernst})} requires, as we can see, that for a condensed solution $K = 1$.
\vspace*{-2mm}
{\color{blue} 
\begin{quote}
Indeed, the last relationship with $r(t)=0 \;(\forall\: t)$, and thus ${\partial\,r\,{\color{blue}(t)}}/{\partial\,t} =0 \; (\forall\: t)$, reduces to $R \:.\: \ln[\:K\,{\color{blue}(T,p)}\:] \,=\,0\:$,
and thus $K=1$ because $R \neq 0$
/ P. Marquet.
\end{quote}
}


The case of thermoneutrality is realized in all its rigor when a body changes into an enantiomorphic form, for example when an optically active molecule is converted into its antipode. Therefore, a mutual solution of the two oppositely active compounds is only in stable equilibrium when it forms a racemic, optically inactive mixture -- a conclusion which is confirmed by the experience that optically active compounds often transform into the racemate on heating. 
The fact that this does not occur in all cases and not even at normal temperature is to be understood as one of the many delay phenomena known in thermodynamics. Other transformations which at least approximately have the character of thermoneutrality were investigated by J. H. van't Hoff in his last work submitted to the Academy of Sciences (the Studies on Synthetic Enzyme Action), and in these too the relationship $K = 1$ was tested and confirmed.


All in all, the Nernst heat theorem {\it\color{blue}(see the footnote \ref{footnote_Planck_Nernst})}, by determining the absolute value of entropy, appears to us as a fundamental complement to the second law of thermodynamics, which allows an exact formulation and therefore an exact test in each individual case. As with the two laws, a single exception would deprive it of its rank, and this is precisely what makes it so important for thermodynamics.

\subsection{\underline{{\color{blue} Planck-Nernst's hypotheses and energy? (p.17-17)}}}
\label{Subsection-I-Planck-Nernst-energy}
\vspace*{-2mm}

One could rightly raise the question as to why what applies to the second law does not also apply to the first, i.e. why the determination of the absolute value of energy, as it is done according to a remark I made above by the principle of relativity, is not accompanied by similar far-reaching consequences for thermodynamics. 

In fact, if the energy of each substance can be measured individually according to its absolute value, then the heat produced by the reaction of any two substances can be calculated directly by subtraction. 
This is certainly indisputable, but unfortunately the principle of this theorem does not correspond to its practical significance. 
In order to be able to utilise the method thermodynamically, one would have to be able to measure the inert mass of a substance down to a millionth of a mg. Then, however, it would have to be shown that water vapour, for example, has a noticeably smaller mass than oxyhydrogen gas at the same temperature.

One should never claim that something that is logically permissible is for all time outside the realm of possibility, but for the time being this consideration serves us only as further proof that the value of a proposition depends not only on its correctness but also on its fertility.

\vspace*{0mm}
\begin{center}
---------------------------------------------------
\end{center}
\vspace*{-7mm}

\section{\underline{II. {\color{blue}Second part: the meaning of Nernst's heat theorem? (p.18-23)}}}
\label{Section-II}
\vspace*{-2mm}

But what is the actual, deeper physical-chemical meaning of Nernst's heat theorem?
{\it\color{blue}(see the footnote \ref{footnote_Planck_Nernst})}

It is probably no longer necessary, least of all in front of an assembly of chemists, to explain further that the question of the atomistic meaning of such a fundamental theorem is a justified and necessary one; and not only because it promises greater clarity, but mainly because it alone can help to uncover laws and connections in the colourful interplay of phenomena that are not touched upon by pure thermodynamics at all. 

Of course, we are entering a new and unfamiliar land of hypotheses right from the start, and if I now attempt to guide you along a viable path, I must confess from the outset that the strange and tantalising prospects that following this path will offer us still require further clarification in detail. 
It seems all the more important to emphasise once again that the question of the validity of Nernst's heat theorem {\it\color{blue}(see the footnote \ref{footnote_Planck_Nernst})} can be treated and answered quite independently of the atomistic hypotheses. 
This is because the purely thermodynamic theory discussed above is completely sufficient for this purpose.

Since the Nernst heat theorem {\it\color{blue}(see the footnote \ref{footnote_Planck_Nernst})} is an entropy theorem, there can be no doubt that its atomistic meaning can only be understood in connection with the atomistic meaning of entropy, i.e. with the atomistic meaning of the second law of thermodynamics. And since the fundamental work of L. Boltzmann, there can be no doubt that in the light of atomism the second law is a probability theorem and entropy is a probability quantity. The thermodynamic theorem (that in every irreversible process the entropy of the bodies involved increases) translated into atomistic language means: through every natural process the bodies involved are transferred to a state of greater probability.
Accordingly, entropy is to be regarded simply as a general measure of probability, and in fact Boltzmann has shown how the entropy of a gas, which is well known from thermodynamics, can be calculated quite independently of all thermodynamics, simply by considering probability, i.e. by applying the elementary theorems of combination theory. One then only has to set the logarithm of the probability of a state proportional to the entropy of this state. 
This simple relationship between entropy and 
probability$\,${\color{blue}\footnote{\label{footnote_Planck_Boltzmann}$\:${\color{blue}Note that it is only Planck who wrote down (in 1900-1901) the formula $S=k_B\:\ln(W)$ when he computed the entropy of radiation, a formula that Boltzmann did not write as such in his book in 1877 (nor later on). Moreover, only Planck} {\color{blue}defined the two constants of nature $a$ and $b$, with $h=b$ and $k_B=b/a$, and then computed their numerical values, even if Max Planck then gave to $k_B$ the name of ``\,Boltzmann's constant\,'' to honour him. / P. Marquet.}}},
assumed to be generally valid, obviously contains the complete explanation of the second law from the standpoint of atomism.



For our present purposes, the additive constant in the expression of entropy, which classical thermodynamics still leaves undetermined, is of particular interest. Because since the Nernst heat theorem {\it\color{blue}(see the footnote \ref{footnote_Planck_Nernst})}, as we have seen, allows the absolute determination of this constant, we may hope to be able to do justice to the atomistic meaning of the Nernst heat theorem if we ask about the atomistic meaning of this constant, i.e. if we try to find the peculiar feature in the atomistic picture used by Boltzmann to calculate entropy that gives rise to the indeterminacy of the additive constant in the expression of entropy.



For this purpose we must go into the method of calculating probability in more detail. For example, let us consider an ideal gas of volume $1$ in a state of a certain total energy, and the calculation of the probability of this state. According to Boltzmann, this can be found in the following way. Represent the total energy of the gas $U$ by a straight line and divide this line into a very large number of equal small sections. Each of these sections then represents a very small energy interval, which extends from a certain value of energy lying between $0$ and $U$ to a value that differs very slightly from this. Then it is obvious that any state of the gas molecules can be illustrated by a numerical image by imagining the molecules as numbered and noting for each energy interval the numbers of those molecules whose energy falls within this interval. The probability sought is then the number of all different numerical images that are possible for the given total energy $U$ of the gas, and the entropy is proportional to the logarithm of the probability.



In the procedure described, every single step is precisely prescribed, with one exception: there is no specification of the size of the energy intervals used. However, it is easy to see that the size of the probability calculated in this way is essentially determined by the number of these small intervals, which serve as a measure for the elementary regions of probability. The smaller these elementary regions are chosen, the larger their number becomes, and the greater the probability sought. If the size of the elementary regions is left undetermined, a certain proportionality constant remains undetermined in the expression of the probability and a certain additive constant remains undetermined in the expression of the entropy, as the logarithm of the probability.



Here we have the answer to the question posed. The additive constant left undetermined by classical thermodynamics in the expression of entropy corresponds, from the atomistic point of view, to the indeterminacy of the elementary regions of probability used for the calculation of entropy; and since the Nernst heat theorem {\it\color{blue}(see the footnote \ref{footnote_Planck_Nernst})} unambiguously determines the value of this constant, the physical content of the Nernst heat theorem {\it\color{blue}(see the footnote \ref{footnote_Planck_Nernst})}, very generally speaking, is that:  \dashuline{the elementary regions of probability are not arbitrarily small, but have a very definite size that can be directly specified in many cases}.

Even if there is hardly a point that can be challenged in the train of thought described above, if one wants to adhere to Boltzmann's connection between entropy and probability at all, it must be said that the result we have arrived at here has something strange and extremely disconcerting for anyone who is closely involved in the study of molecular processes. 
Because the previous atomistic theories do not offer the slightest clue for the delimitation of very specific elementary areas of probability.
Indeed, whether such a delimitation can be given any physical meaning at all seems at first sight to be quite doubtful, if not impossible. 
And the question would probably be whether it is not all too risky to proceed even further along the new path if, strangely enough, investigations of a completely different kind from another side did not lead to the same path.



Here we see once again the case that occurs not infrequently in the history of science, that a new idea capable of development sprouts up almost simultaneously in different places that have no connection with each other, and that its development takes place independently for a time in very different forms in different places, until finally the recognition of its unity breaks through everywhere. 
Once the fusion has been achieved, the different methods are able to lend each other their impetus and thereby at the same time strengthen their own effectiveness.



In the theory of thermal radiation, the blatant contradiction between the radiation formula of classical dynamics and the results of measurements had also led to the strange conclusion that there were very specific elementary regions of probability for radiant heat, and comparison with observations had even made it possible to calculate the size of these elementary regions, the universal quantum of action, with considerable accuracy. Even if one were inclined to consider this coincidence to be a mere coincidence, it must at least appear interesting to compare these results obtained in completely different areas. This has indeed been done, and with a result that could not have been better expected given the diversity of the objects.

Firstly, by comparing the laws of radiation with the laws of gas, a method was found for calculating the elementary quanta of matter and electricity that rivalled the finest direct measurements in terms of accuracy. Should this correspondence also be based only on coincidence? 
But even more, A. Einstein on the one hand, W. Nernst and F. A. Lindemann on the other, have found that with the help of the universal quantum of action the specific heat of a whole series of solid bodies can be calculated, both absolutely and in their dependence on temperature, if certain molecular eigenvibrations are ascribed to the bodies, 
and the eigenvibrations calculated in this way agree with the eigenvibrations optically measured by H. Rubens and H. Hollnagel for some substances (such as \ch{NaCl}, \ch{KCl}, \ch{KBr}), with all the accuracy one could wish for.



In view of these results, it hardly seems permissible to speak of a coincidence here. But whatever one may think about this, it now turns out to be a task that is as important as it is tempting to investigate further the hypothesis that certain very specific elementary regions exist for thermodynamic probability --for that is how I would like to summarize the actual content of the so-called \dashuline{quantum hypothesis}-- and to search for its deeper physical meaning.

This is now an extremely difficult problem, because as simple as it is in many cases to calculate the elementary areas of probability -- O. Sackur recently carried out such a calculation for gases -- the question of their physical-chemical origin is just as complicated. 
It must be borne in mind that the task here is to draw conclusions from a merely statistical law to the dynamic law, i.e. to the causal relationship between the individual processes. 
This is a task of a similar nature to that of deducing the chemical forces acting between the reacting molecules from the speed of a chemical reaction. 
Naturally, there are a whole range of possibilities here from the outset, and it is not surprising that the views of the various researchers still differ widely today.

The simplest, most naive explanation, so to speak, would be that energy itself has an atomistic structure. Then, of course, the existence of certain discrete elementary areas of probability would be explained without further ado. But there can be no question of realising such a view for the simple reason that we cannot conceive of the kinetic energy of a straight-line progressive motion as being discontinuously variable. On the other hand, it is often argued that the energy of electromagnetic wave radiation, or at least the vibrational energy of electrons, which is also the source of thermal radiation, is atomistically constituted, in that it should always be a whole multiple of a certain quantum of energy. I myself used to subscribe to the latter assumption, but have now retreated from it because I still consider it too radical to be acceptable in all
cases$\,${\color{blue}\footnote{\label{footnote_Planck_quantum}$\:${\color{blue}It took a long time for Max Planck to accept, after more than ten years, and thus after 1912 and after having exhausted his strength to prove the contrary, that the quanta of energy that he introduced in 1899-1900-1901 to establish his formula for the energy of black bodies had to have a meaning against all physical intuition. / P. Marquet.}}}.
However, there is no need to go that far. The quantum hypothesis only requires that certain discontinuities lie hidden in the elementary laws that govern the atomistic forces, from which the discrete areas of probability then result. Nothing can be said from the outset about the nature of these discontinuities. In particular, it should be noted that the quantum-like structure does not initially relate to energy, but to probability. One can only speak of energy quanta in the case of periodic processes. In my opinion {\it\color{blue}(see the footnote \ref{footnote_Planck_quantum})}, the quantum hypothesis can be fully taken into account if only the emission of energy in a periodically oscillating molecular oscillator is regarded as quantum, whereas absorption, at least in the case of radiant heat, is regarded as completely continuous. For non-periodic processes, A. Sommerfeld recently outlined the main features of a very bold and very interesting quantum theory, in which, of course, only quanta of action, not quanta of energy, play a role.

This dazzling diversity of views should not be interpreted to the detriment of the quantum hypothesis itself. On the contrary, it is precisely when as many directions as possible are explored, when each researcher, unperturbed by objections that he does not consider valid, continues to pursue his own path in areas in which he himself feels most confident, that we can first hope that the true character of the hypothesis will be revealed. 
In fact, in addition to thermal radiation and specific heat, a large number of other physical processes have gradually been related to the quantum hypothesis: the Doppler effect in canal rays, the photoelectric effect, the ionisation voltage, the generation of X-rays and gamma rays and their inversion (the triggering of secondary cathode rays by X-rays), the electrical conduction resistance, the thermoelectric forces, the law of formation of the spectral series lines, the electron emission in chemical reactions -- everywhere one can, at least with some good will, get on the track of the still very mysterious workings of the universal quantum of action. 
Indeed, the strange fact established by O. Hahn and his co-workers that a radioactive substance, if it is only chemically uniform, emits $\beta$ rays of quite definite velocities, sees the quantum emission ad ocul, so to speak.



Of course, most of the work still remains to be done, and some seemingly promising findings will probably still fall off the tree of knowledge as deaf blossoms.
But a start has been made: the quantum hypothesis will no longer disappear from the world. 
The laws of thermal radiation will ensure that.
And I believe I am not going too far when I express the opinion that this hypothesis lays the foundation for the construction of a theory that will one day be destined to shed new light on the details of the fine, rapid processes of the molecular world.

\vspace*{-2mm}
\begin{center}
---------------------------------------------------
\end{center}
\vspace*{-3mm}

\newpage
\section{\underline{\color{blue} The German Paper of Planck (1912)}}
\label{Section-PDF-1912}
\vspace*{-6mm}
\vspace*{9mm}

\begin{figure}[hbt]
\centering
\includegraphics[width=0.66\linewidth]{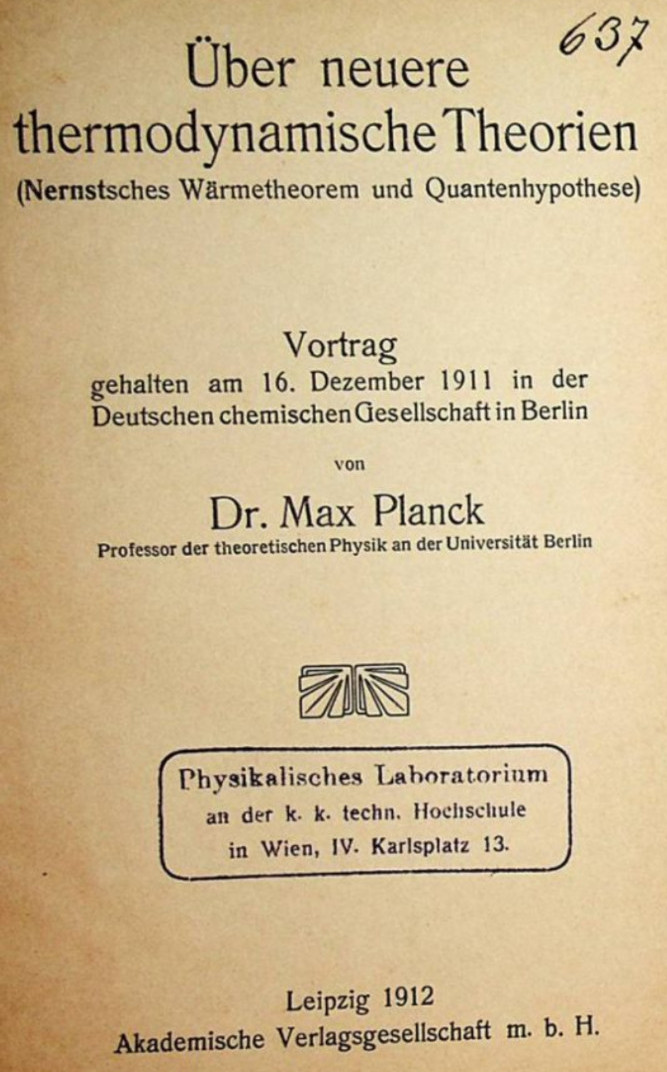}
\end{figure}
\begin{figure}[hbt]
\centering
\includegraphics[width=0.75\linewidth]{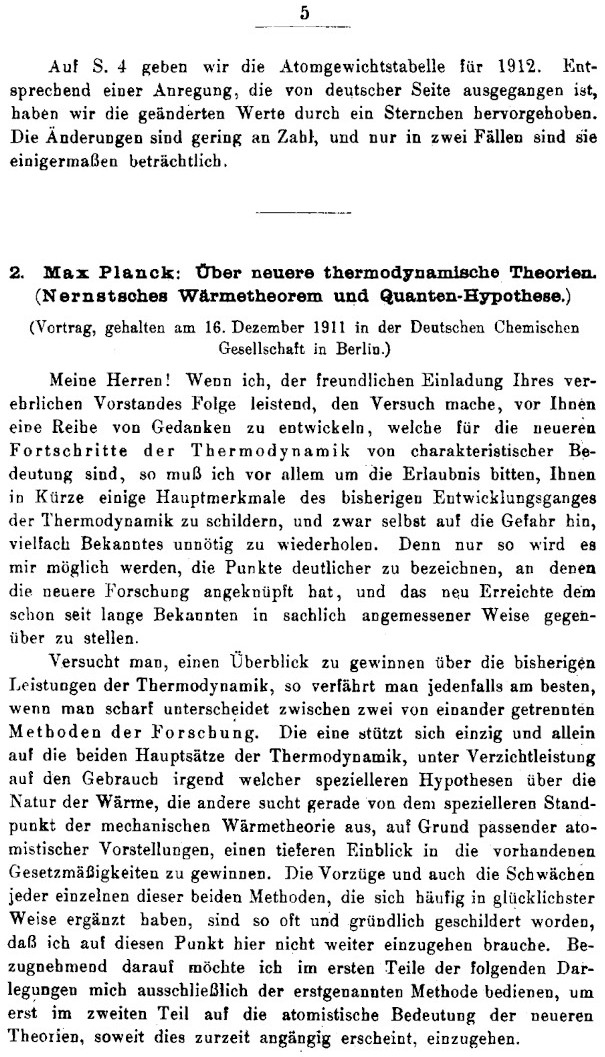}
\end{figure}
\begin{figure}[hbt]
\centering
\includegraphics[width=0.75\linewidth]{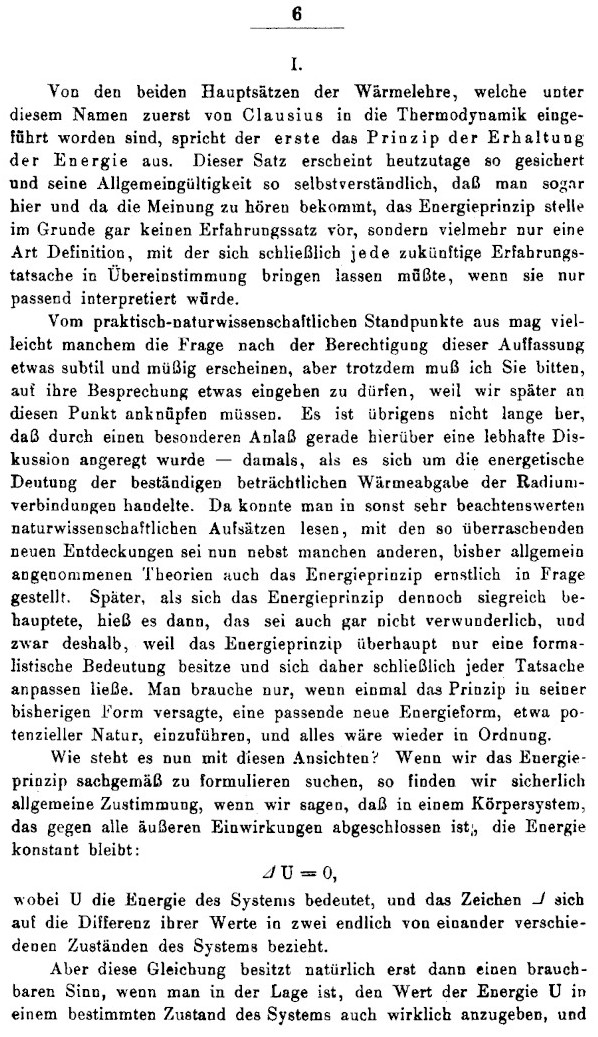}
\end{figure}
\begin{figure}[hbt]
\centering
\includegraphics[width=0.75\linewidth]{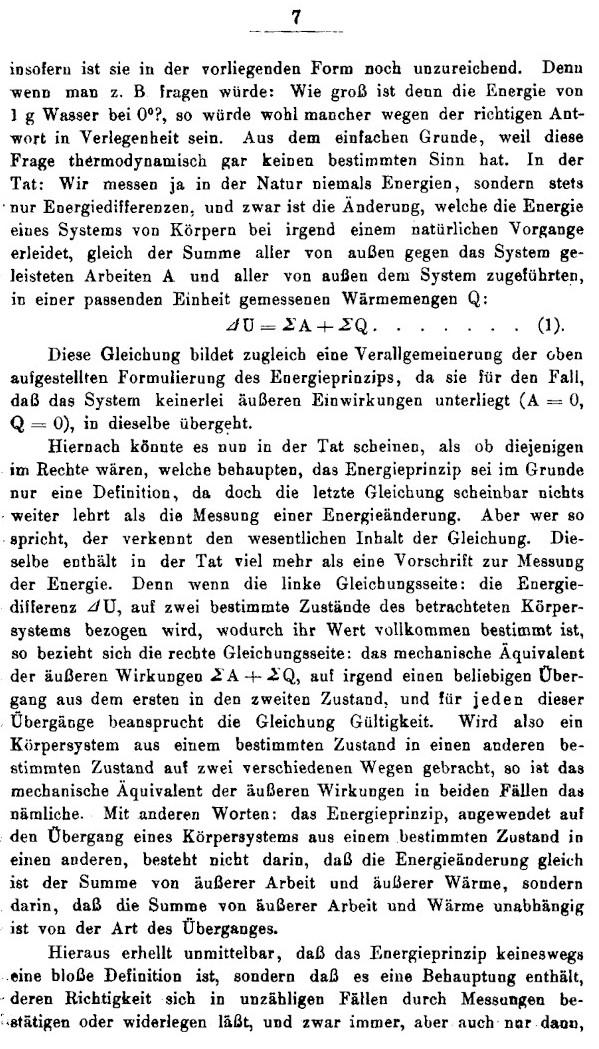}
\end{figure}
\begin{figure}[hbt]
\centering
\includegraphics[width=0.75\linewidth]{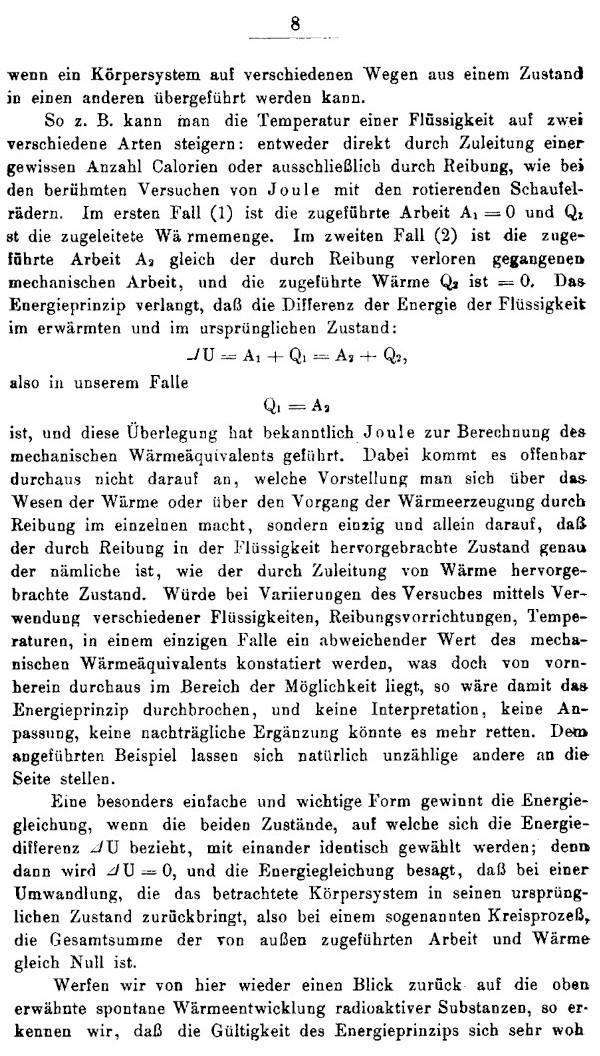}
\end{figure}
\begin{figure}[hbt]
\centering
\includegraphics[width=0.75\linewidth]{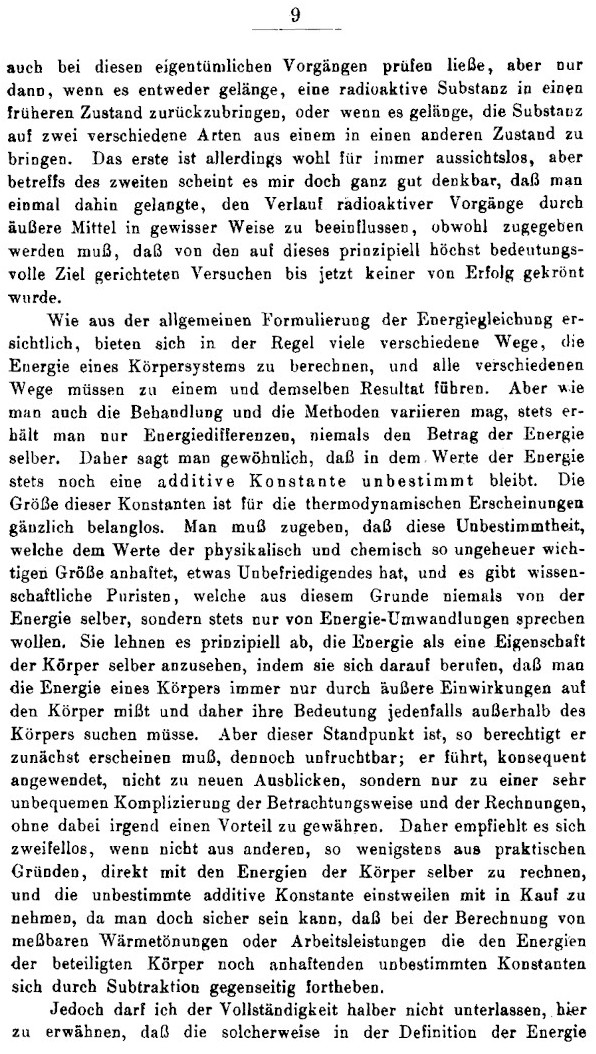}
\end{figure}
\begin{figure}[hbt]
\centering
\includegraphics[width=0.75\linewidth]{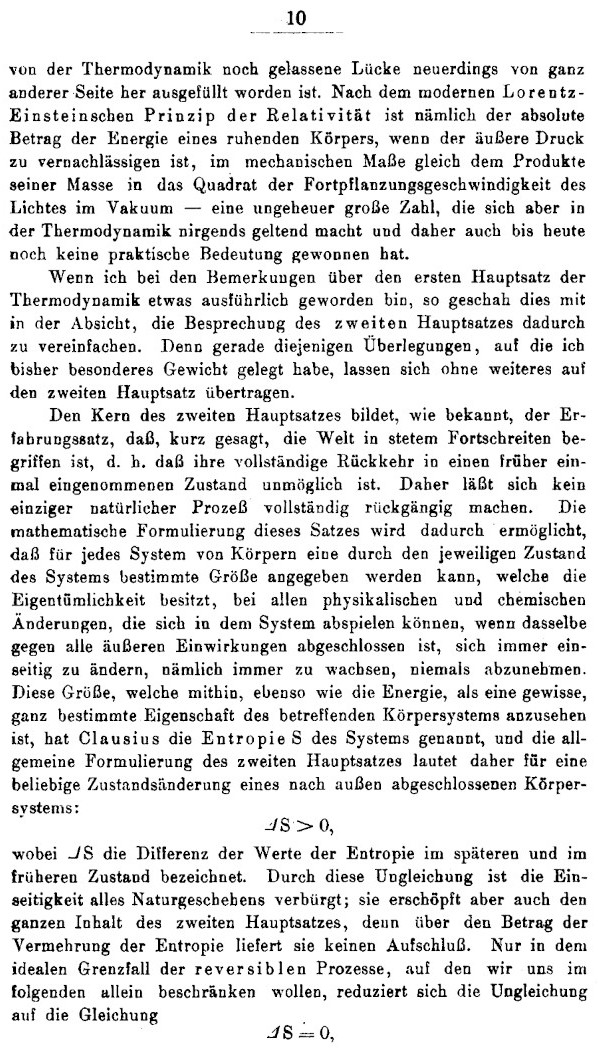}
\end{figure}
\begin{figure}[hbt]
\centering
\includegraphics[width=0.75\linewidth]{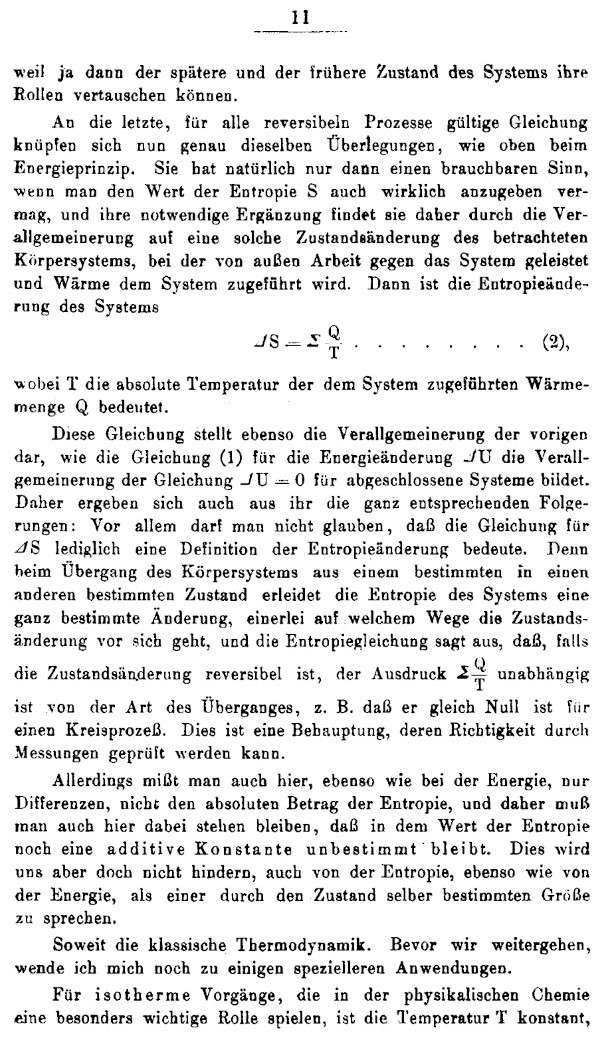}
\end{figure}
\begin{figure}[hbt]
\centering
\includegraphics[width=0.75\linewidth]{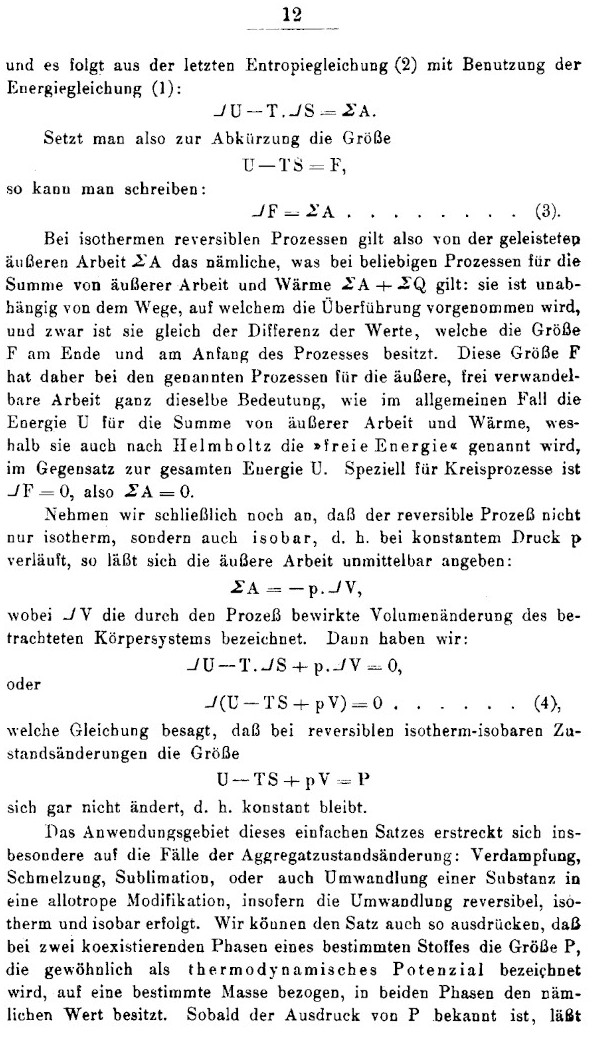}
\end{figure}
\begin{figure}[hbt]
\centering
\includegraphics[width=0.75\linewidth]{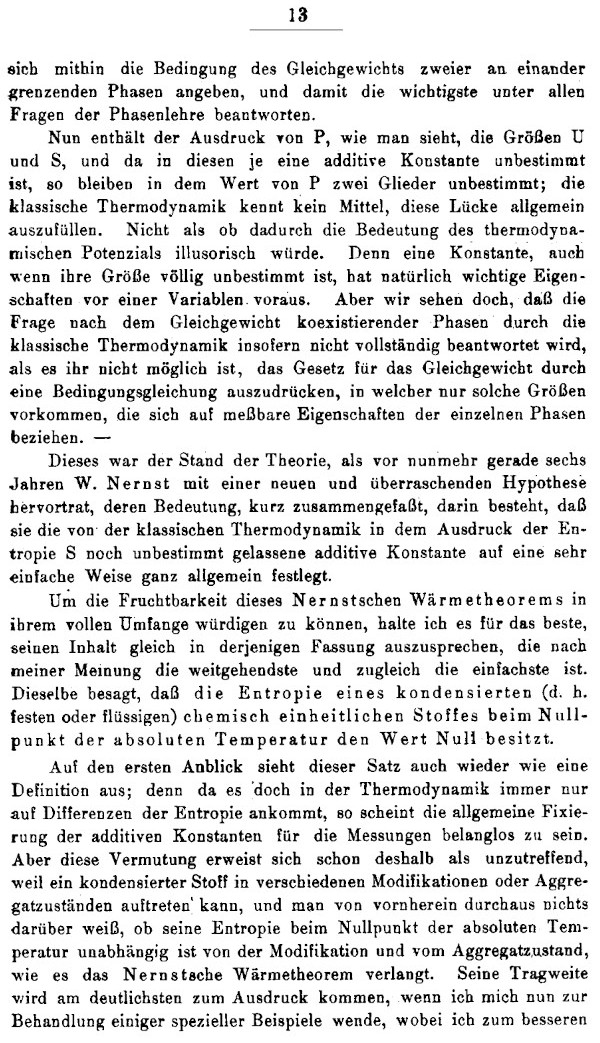}
\end{figure}
\begin{figure}[hbt]
\centering
\includegraphics[width=0.75\linewidth]{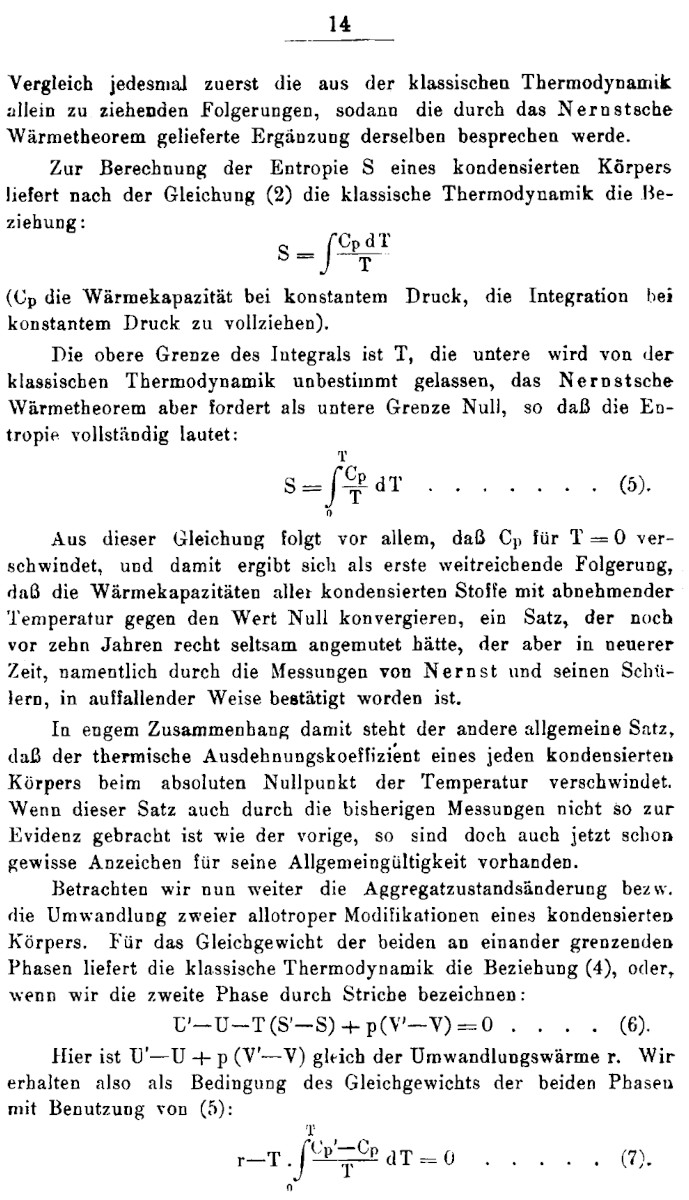}
\end{figure}
\begin{figure}[hbt]
\centering
\includegraphics[width=0.75\linewidth]{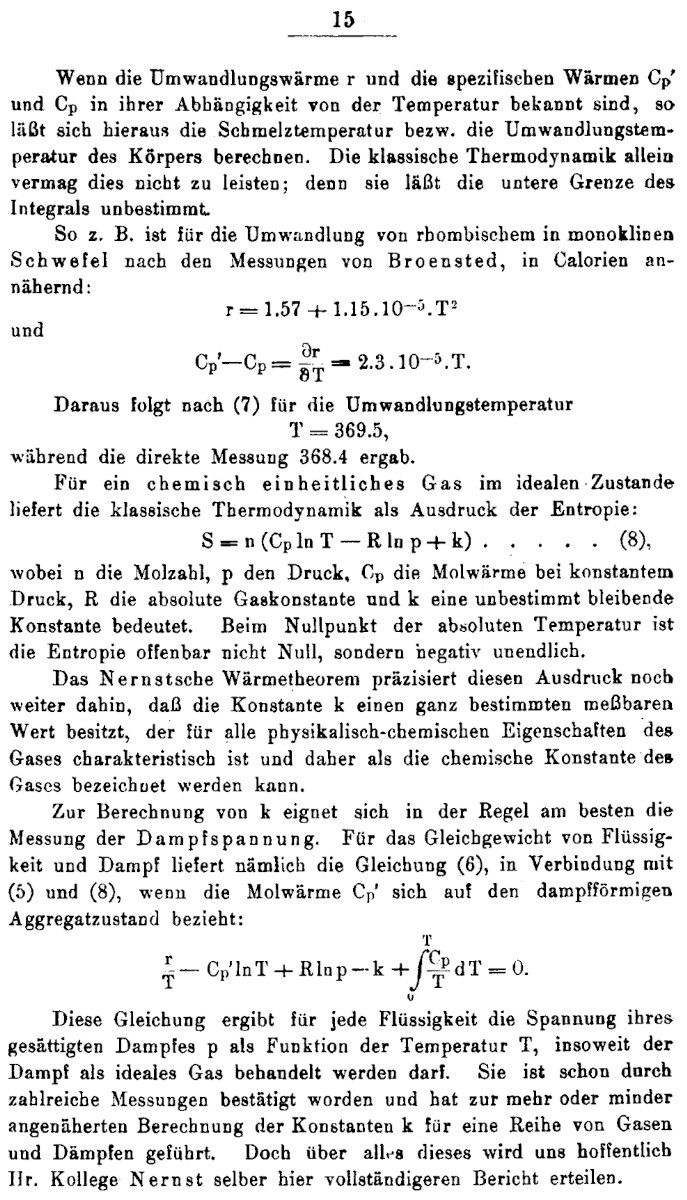}
\end{figure}
\begin{figure}[hbt]
\centering
\includegraphics[width=0.75\linewidth]{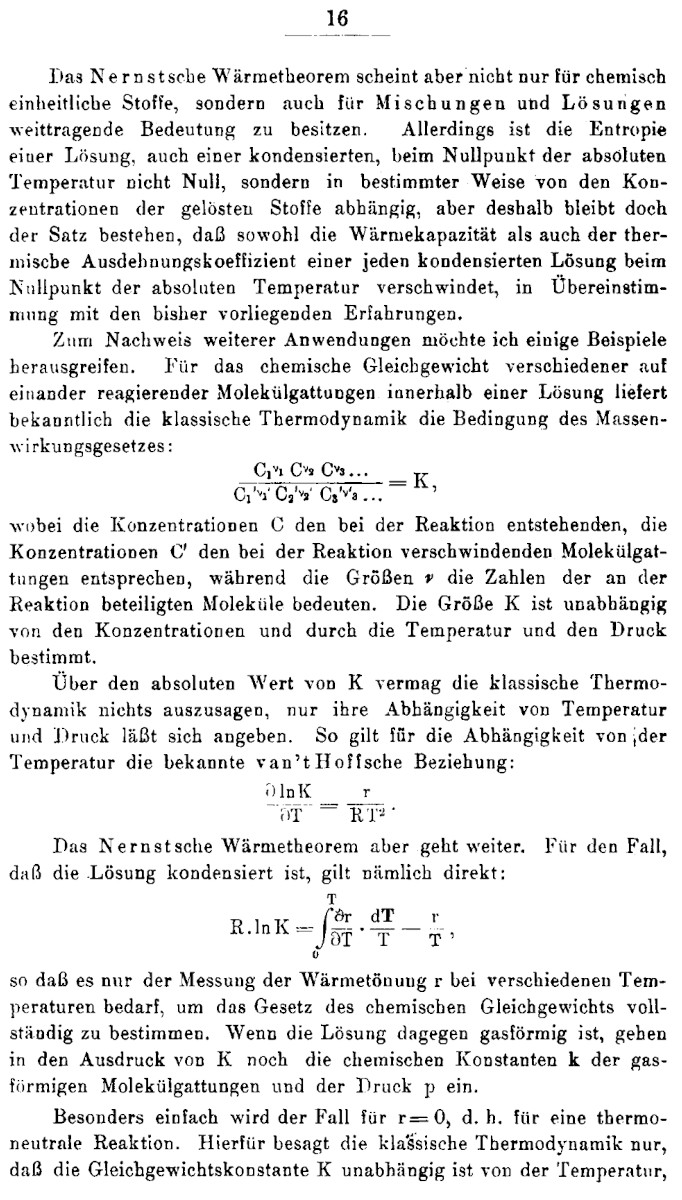}
\end{figure}
\begin{figure}[hbt]
\centering
\includegraphics[width=0.75\linewidth]{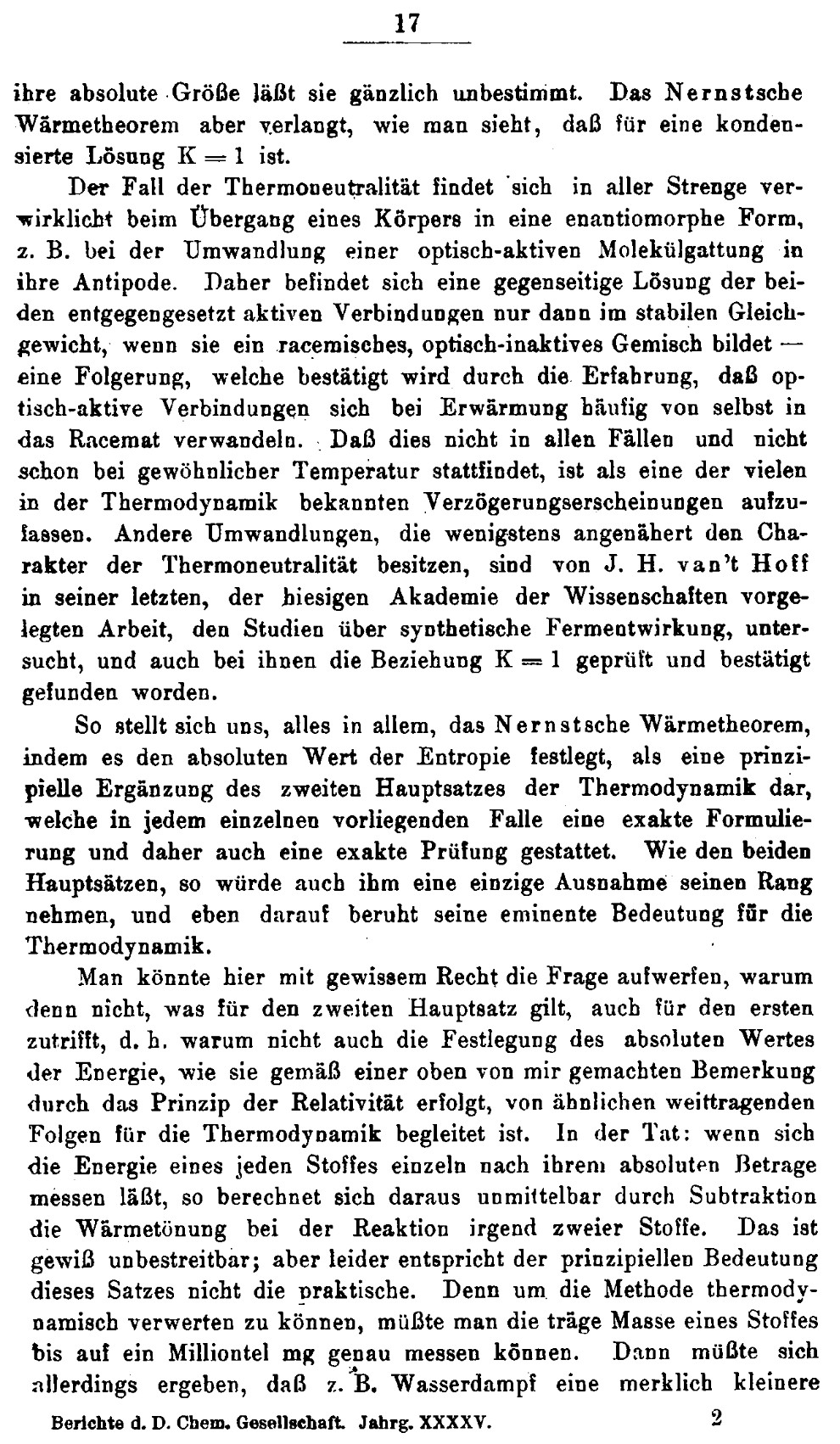}
\end{figure}
\begin{figure}[hbt]
\centering
\includegraphics[width=0.75\linewidth]{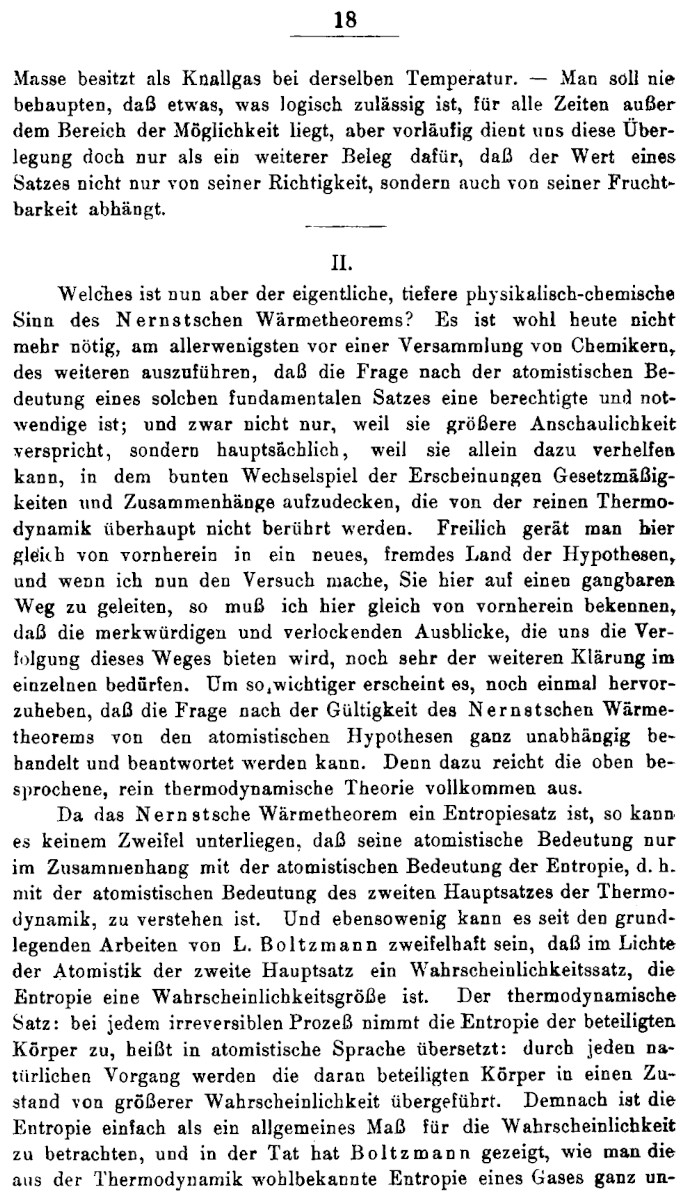}
\end{figure}
\begin{figure}[hbt]
\centering
\includegraphics[width=0.75\linewidth]{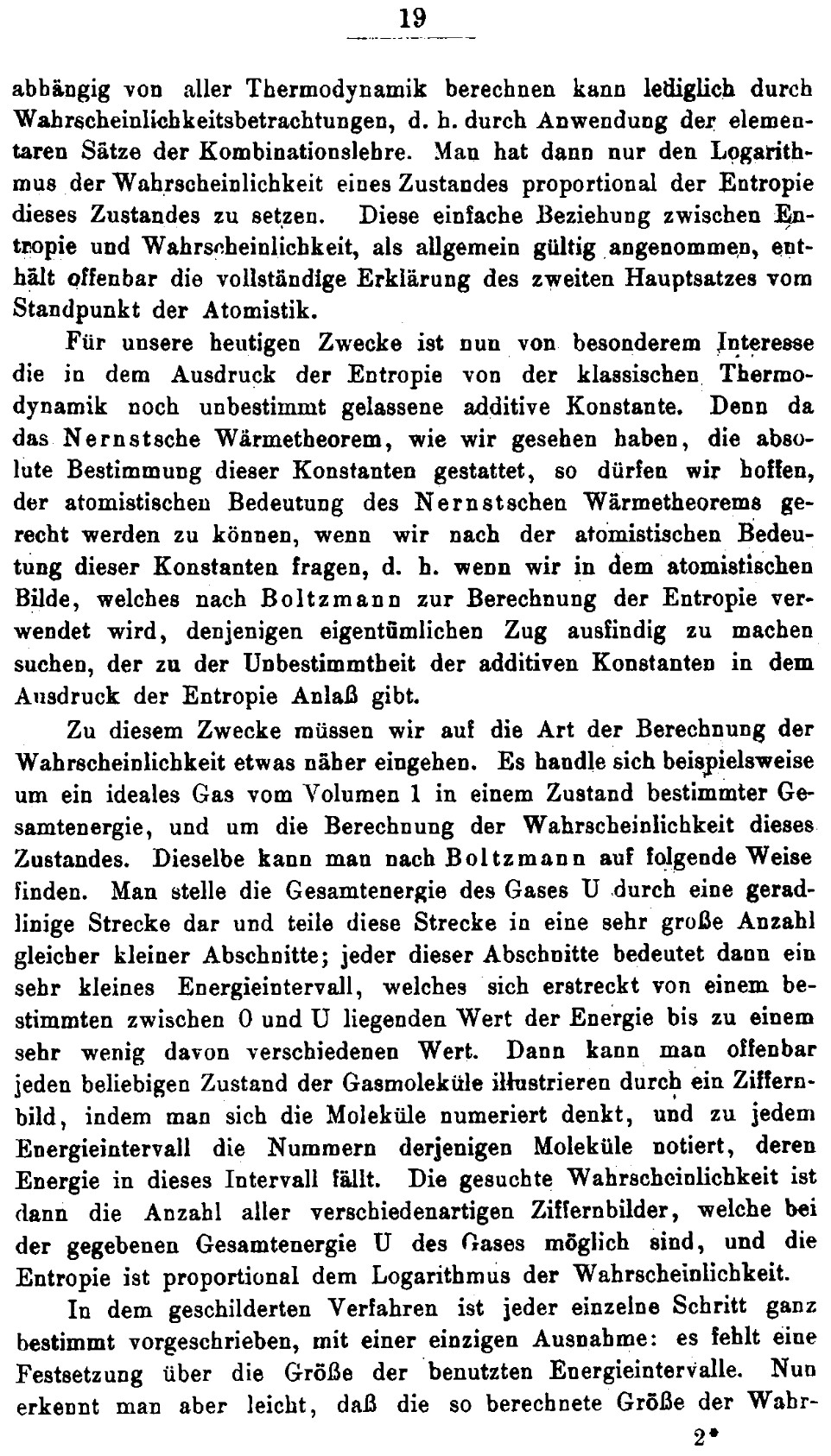}
\end{figure}
\begin{figure}[hbt]
\centering
\includegraphics[width=0.75\linewidth]{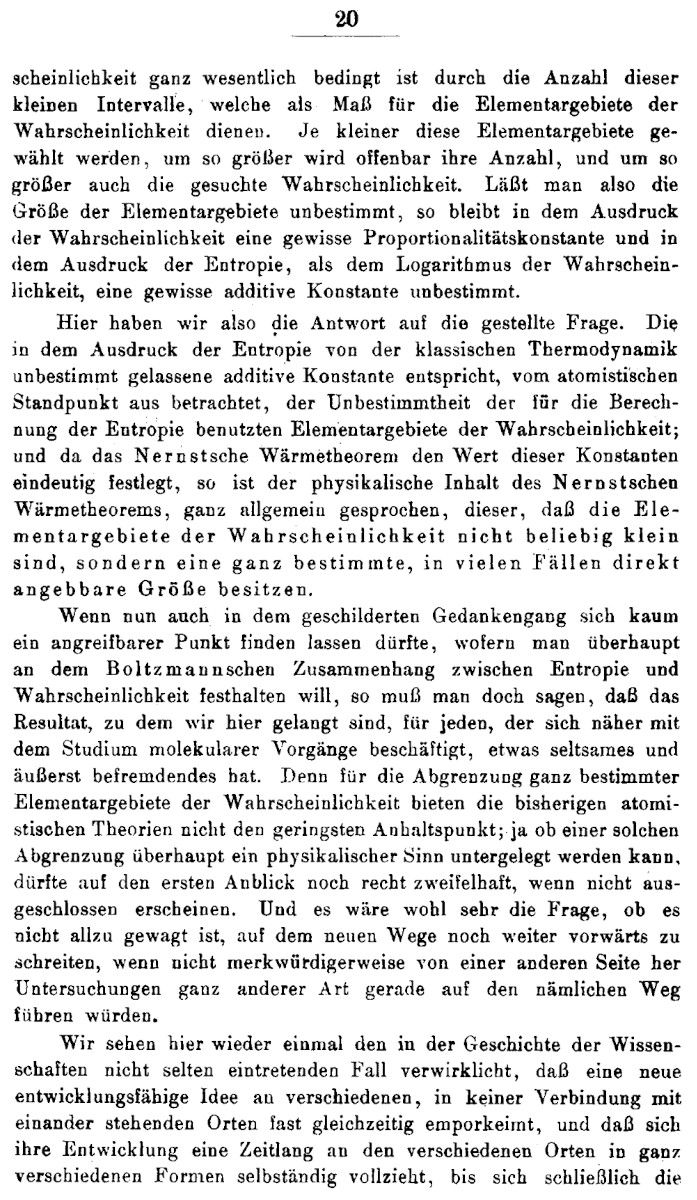}
\end{figure}
\begin{figure}[hbt]
\centering
\includegraphics[width=0.75\linewidth]{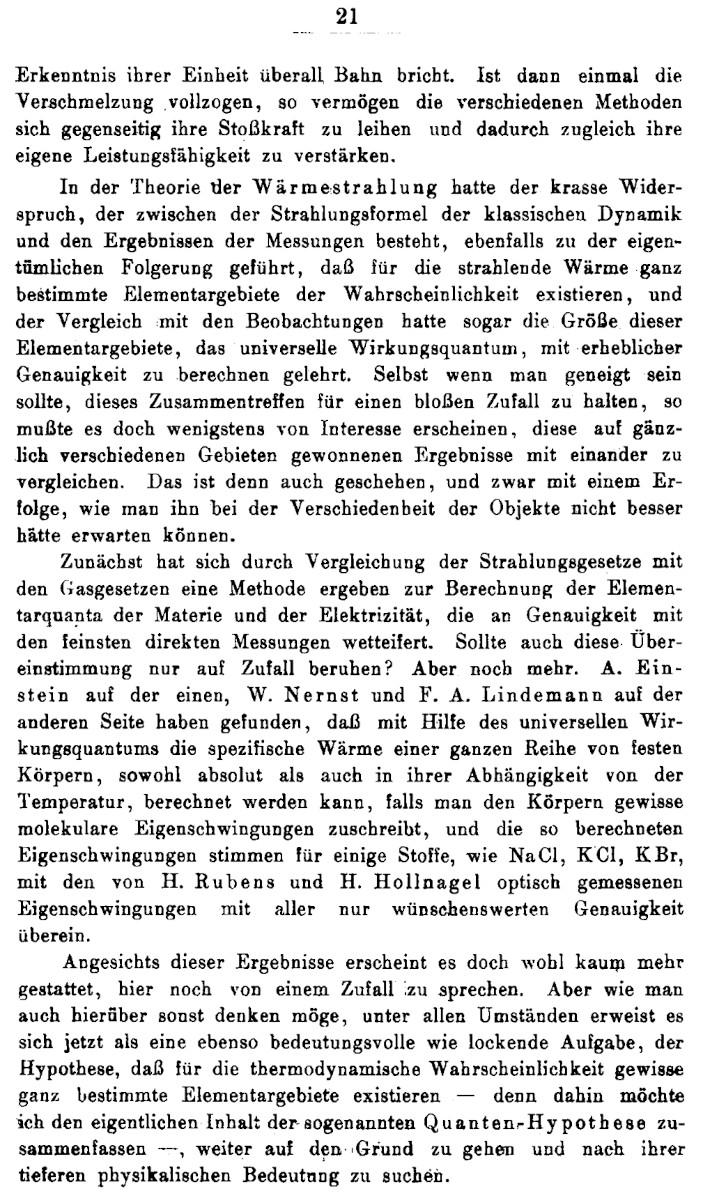}
\end{figure}
\begin{figure}[hbt]
\centering
\includegraphics[width=0.75\linewidth]{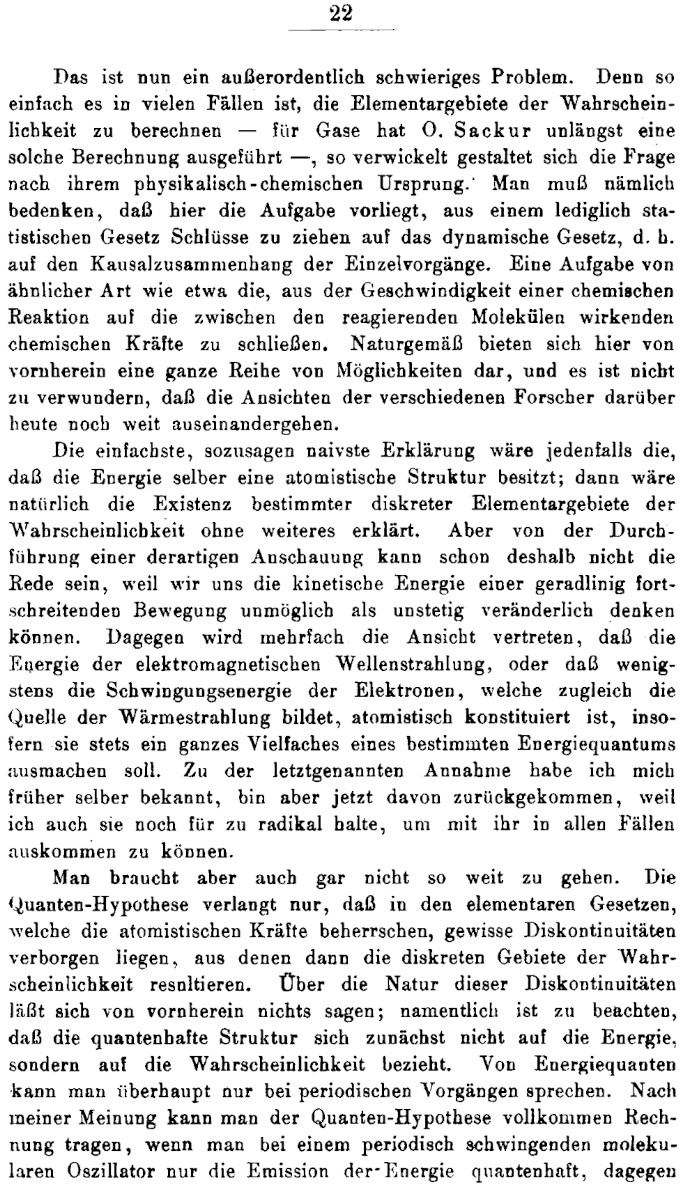}
\end{figure}
\begin{figure}[hbt]
\centering
\includegraphics[width=0.75\linewidth]{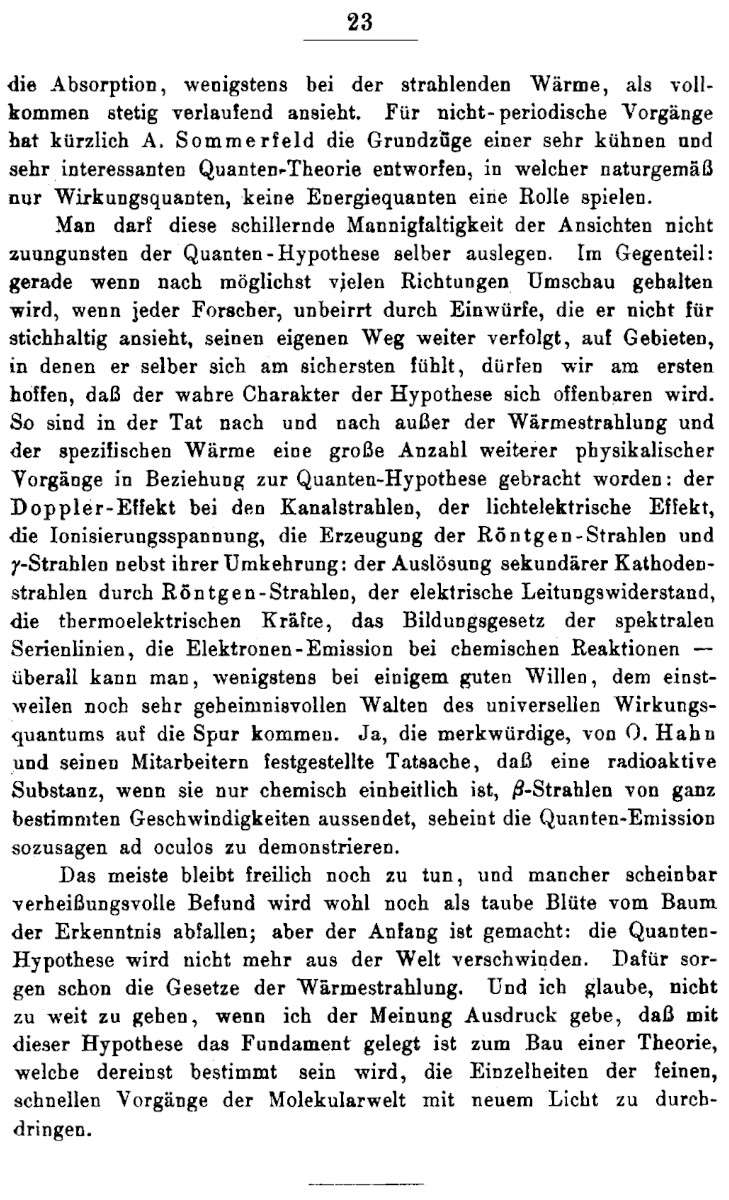}
\end{figure}

\end{document}